\documentclass{spie}

\usepackage{graphicx}

\title{\Large \bf Quantizing Knots and Beyond}

\author{Louis H. Kauffman\supit{a} and Samuel J. Lomonaco Jr.\supit{b}
\skiplinehalf
\supit{a} Department of Mathematics, Statistics and Computer Science  
(m/c 249), 851 South Morgan Street, University of Illinois at Chicago,
Chicago, Illinois 60607-7045, USA \\
\supit{b} Department of Computer Science and Electrical Engineering, University of
Maryland Baltimore County, 1000 Hilltop Circle, Baltimore, MD 21250, USA}
 
\authorinfo{Further author information: L.H.K.  E-mail: kauffman@uic.edu,
S.J.L. Jr.: E-mail: lomonaco@umbc.edu}
  
\begin{document} 

\newcommand{\Across}{\raisebox{-0.25\height}{\includegraphics[width=0.5cm]{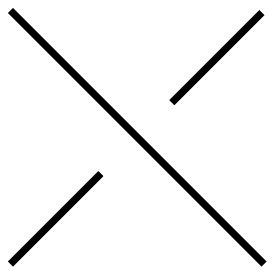}}}
\newcommand{\Bcross}{\raisebox{-0.25\height}{\includegraphics[width=0.5cm]{B.eps}}}
\newcommand{\Asmooth}{\raisebox{-0.25\height}{\includegraphics[width=0.5cm]{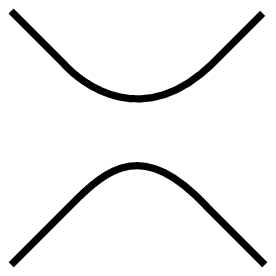}}}
\newcommand{\Bsmooth}{\raisebox{-0.25\height}{\includegraphics[width=0.5cm]{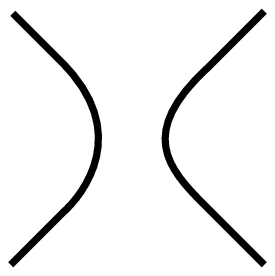}}}
\newcommand{\Rcurl}{\raisebox{-0.25\height}{\includegraphics[width=0.5cm]{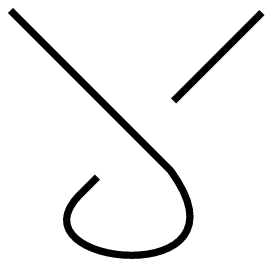}}}
\newcommand{\Lcurl}{\raisebox{-0.25\height}{\includegraphics[width=0.5cm]{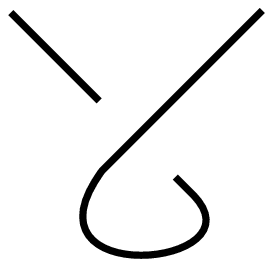}}}
\newcommand{\Arc}{\raisebox{-0.25\height}{\includegraphics[width=0.5cm]{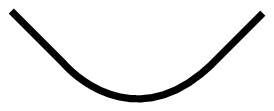}}}

 \maketitle

\begin{abstract}
This paper formulates a generalization of our work on quantum knots to explain how
to make quantum versions of algebraic, combinatorial and topological structures. We include a description of previous work on the construction of Hilbert spaces from the states of the bracket polynomial with applications to algorithms for the Jones polynomial and relations with Khovanov homology. The purpose of this paper is to place such constructions in a general context of the quantization of mathematical structures.
\end{abstract}


\section{Introduction}
 The purpose of this paper is to show some of the uses and definitions for making quantum 
 information versions of combinatorial and topological structures. We begin in Section 2 by reviewing
 our previous work on quantum knots where we model the topological information in a knot by a state
 vector in a Hilbert space that is directly constructed from mosaic diagrams for the knots. In Section 3 we discuss some issues for these models relating topological and quantum entanglement. In Section 4 we
 give a general definition of quantization of mathematical structures and apply it to the quantization of 
 the set of classical knots (embeddings of a circle into three dimensional space) and the group of 
 homeomorphisms of three dimensional space that acts on this set of embeddings. The Hilbert space that results from this set of embeddings is very large, but descriptive of the sort of knotting phenomena that may occur in nature such as knotted vortices in super-cooled Helium or knotted gluon fields.
 In Section 5 we define quantum Gauss codes and show how to formulate quantum knot theory in the context of these codes. In Section 6 we define a quantum version of directed graphs (see also 
 \cite{LomQknots1,LomQknots2}) and in Section 7 we define 
 a quantization of the words in a group presentation. In the remaining sections of the paper we outline previous work  \cite{KhoJones} defining Hilbert spaces that correspond to enhanced states of the bracket state sum for the Jones polynomial. These constructions lead to quantum algorithms for the Jones polynomial that are of conceptual interest because of their relationship with Khovanov homology. 
 In this last part of the paper we are quantizing an unusual combinatorial structure - the set of enhanced states of the bracket polynomial. By making this set of states into a Hilbert space, we combine our notions of quantization with Khovanov's construction that associates a chain complex to a knot diagram. This paper contains a wide range of constructions and it is intended to provide a springboard for discussion of the use of such quantization in quantum information theory. For an in depth view of basic aspects of our quantization procedures the reader is referred to \cite{LomQknots2,KauffQknots1}. The 
 present paper is an expanded version of a paper \cite{KauffQknots1} presented at the Spie Conference in Orlando, Florida in April 2011.
 \bigbreak

\section{Mosaic Quantum Knots}
We begin by explaining the basic idea of mosaic quantum knots as it appears  in 
our papers \cite{KauffQKnots,LomQKnots,LomQknots1}. A recent application
of quantum knots can be found \cite{Shor} in a recent paper by Farhi, Gosset, Hassidim, Lutominski and 
Shor. There the reader will find proof that quantum knots is a money-making idea.
\bigbreak

View Figures 1, and 2. In the left-most part of Figure 1 we illustrate a mosaic version of a trefoil knot using a $4 \times 4$ space of tiles. In Figure 2 we show the eleven basic 
tiles that can be repeated used in $n \times n$ tile spaces to make diagrams for any classical
knot or link. So far this is a method for depicting knots and links and has no quantum interpretation.
However, as in our previous papers, we use the philosophy that given a well-defined discrete set of 
objects, one can define a vector space with an orhonormal basis that is in one-to-one correspondence
with these objects. Here we let $V$ be the complex vector space with basis in one-to-one 
correspondence with the set of eleven basic tiles shown in Figure 2. An $n \times n$ mosaic as shown in Figure 1 is then regarded as an element in the tensor product of $n^{2}$ copies of $V$.
We order the tensor product by consectively going through the rows of the mosaice from left to right 
and from top to bottom. In this way, knot diagrams represented by $n \times n$ mosaics are
realized as vectors in $H_{n} = V \otimes V \otimes \cdots \otimes V$ where there are 
$n^{2}$ factors in this tensor product.
\bigbreak

Isotopy moves on the mosaic diagrams are encoded by tile replacements that induce unitary transformations on the Hilbert space. We refer to \cite{LomQKnots} for the details. The upshot of this fomulation of 
isotopies of the knots is that the diagrammatic isotopies correspond to unitary transformations of the 
Hilbert space $H_{n}$ when the isotopies are restricted to the $n \times n$ lattice. In this way we obtain for each $n$ a group of isotopies $A_{n}$ that we call the {\em ambient  group}. This has the advantage that it turns a version of the Reidemeister moves on knot and link diagrams into a group
and it provides for a quantum formulation not just for the knot and link diagrams, but also for their 
isotopies. Knots and links are usually regarded as entities of a classical nature. By 
making them into quantum information, we have created a domain of quantum knots and links.
\bigbreak

There are many problems and many avenues available for the exploration of quantum knots and links.
It is not the purpose of this paper to specialize in this topic. We show them in order to emphasize the idea that one can quantize combinatorial categories by formulating appropriate Hilbert spaces for their
objects and morphisms. However, it is worth mentioning that other diagrammatic categories for knots and links can be easily accomodated in the mosaic link framework. View Figure 1 again and examine the middle and right diagrams in the figure. Here we show diagrams containing white and black graphical nodes. These can be interpreted for extensions of knot theory to virtual knots or knotted graph 
embeddings. One extends the vector space for the basic tiles and then also adds moves that are
appropriate for the theory in question. In the case of virtual knot theory and the theory of knotted graphs, there is no problem in making these extensions. We will carry them out in a separate paper. The point of 
this section has been to remind the reader of our previous work, and to point to these avenues along which it can be extended. The reader will find an in depth treatment of many aspects of these quantization procedures in \cite{LomQknots2}. After a a discussion of ideas and examples, the present paper concentrates on a quantization related to Khovanov homology in sections 7, 8 and 9.
\bigbreak

Lets go back to classical knot theory in mosaic form. One of the problems in studying this theory is the 
matter of articulating invariants of knots so that they are quantum observables for the theory.
Many invariants such as the Jones polynomial and even the bracket model for the Jones polynomial (which we will discuss later in this paper) seem to be resistant to formulation as quantum 
observables. However, there is a very general result for mosaic quantum knots that is intellectually satisfying that we have proved in our earlier work \cite{LomQKnots}.
\bigbreak

\noindent {\bf Theorem 1.} Let $| K \rangle$ be a mosaic knot diagram in an $n \times n$ lattice.
Then there is a quantum observable $\chi(K)$ such that $\chi(K) | K' \rangle = | K' \rangle $ if and only if
$ | K' \rangle $ is in the orbit of $ | K \rangle $ under the action of the ambient  group
$A_{n}.$ When $|K' \rangle$ is not in the orbit, then $\chi(K) | K' \rangle = 0.$ In other words $\chi(K)$ is a characteristic function for the knot-type of $K$ in the $n \times n$
mosaic lattice.
\bigbreak

\noindent {\bf Proof.} Define $\chi(K)$ by the formula
$$\chi(K) = \sum_{|K' \rangle \in Orbit(K)} | K' \rangle \langle K' |$$
where $Orbit(K)$ denotes the orbit of $| K \rangle$ under the action of the ambient 
group $A_{n}.$ Note that $Orbit(K)$ is a finite set. The Theorem follows directly from this 
definition. //
\bigbreak

This Theorem does not make invariants of knots that are efficient to calculate, but it is intellectually 
satisfying to know that, in principle, in the $n \times n$ lattice, we can distinguish two diagrams that
are inequivalent by the Reidemeister moves for that lattice size. Furthermore, we can use the 
characteristic observables $\chi(K).$  to make observables for any real valued classical knot invariant.
By a classical knot invariant, we mean a function on standard knot diagrams that is invariant under the
usual graphical Reidemeister moves. Such a function is also tautologically defined on mosaic diagrams
and is invariant under the mosaic moves for any $n \times n$ mosaic lattice. For example, the 
Jones polynomial \cite{JO} $V_{K}(t)$ is a Laurent polynomial valued invariant. 
By taking the variable $t$
 in the Jones polynomial to be a specific real number, we obtain from the Jones polynomial, a real-valued classical knot invariant.
\bigbreak

\noindent {\bf Theorem 2.}  Let $Inv(K)$ denote a real-valued classical invariant of knots and links.
Then there is an observable $O$ on the Hilbert space for any $n \times n$ mosaic lattice such 
that $O|K \rangle = Inv(K)|K \rangle$ for any knot vector $| K \rangle$ in the lattice. In this sense,
any real-valued classical knot invariant corresponds to a quantum observable whose eigenvalues
are the values of this invariant.
\bigbreak

\noindent {\bf Proof.} Define the observable $O$ by the formula
$$O = \sum_{K} Inv(K) \chi(K)$$ where $K$ runs over one representative for each ambient group  orbit in  the $n \times n$ mosaic lattice. Here $\chi(K)$ is the observable defined in Theorem 1. The 
Theorem then follows directly from this definition. //
\bigbreak

In this sense, the quantum observables for mosaic quantum knots are universal with respect to 
real-valued classical knot invariants. It remains to be seen if there are such observables that have
a better than classical efficiency of calculation.
\bigbreak

Other issues for quantum knots involve considering superpositions of them and properties of these
superpositions. We refer the reader to \cite{KauffQKnots,LomQKnots,LomQknots1} for examples along these lines.
\bigbreak

 \begin{figure}[htb]
     \begin{center}
     \begin{tabular}{c}
$ 
\begin{array}{cccc}
\includegraphics[width=1cm]{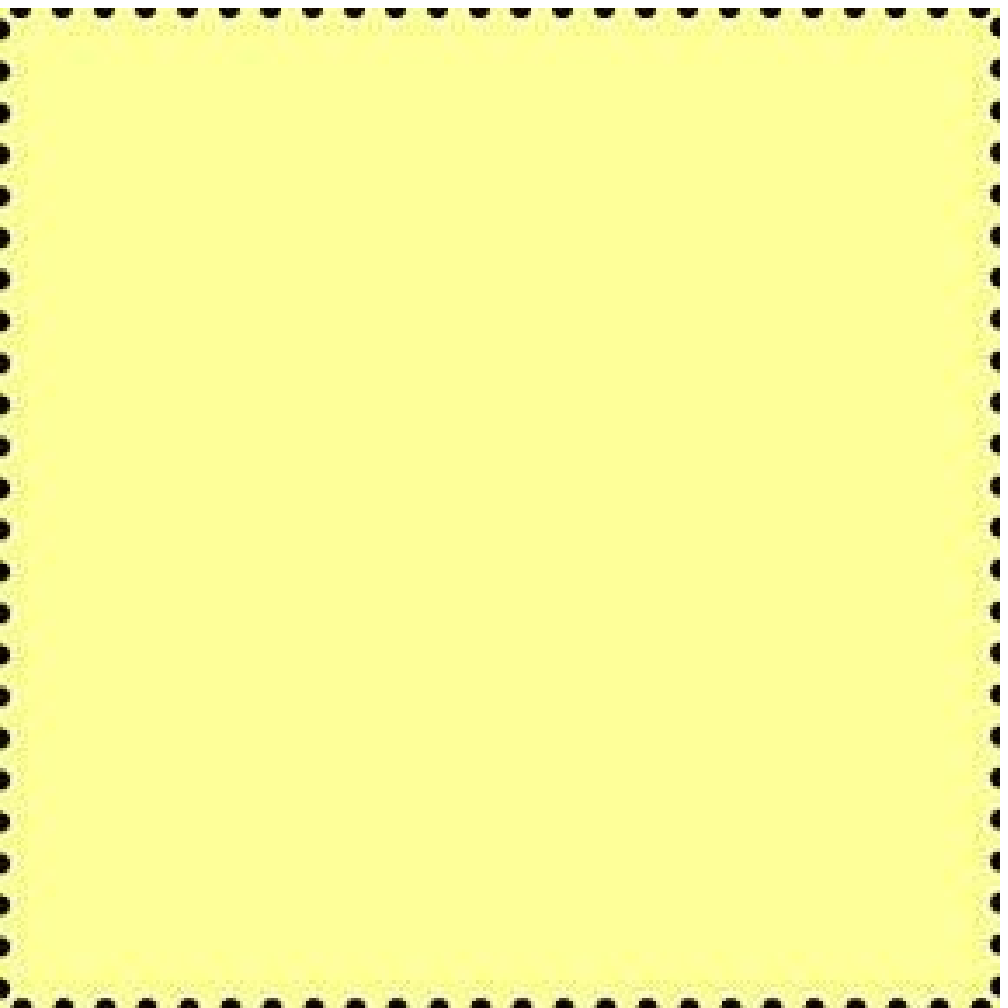} & \includegraphics[width=1cm]{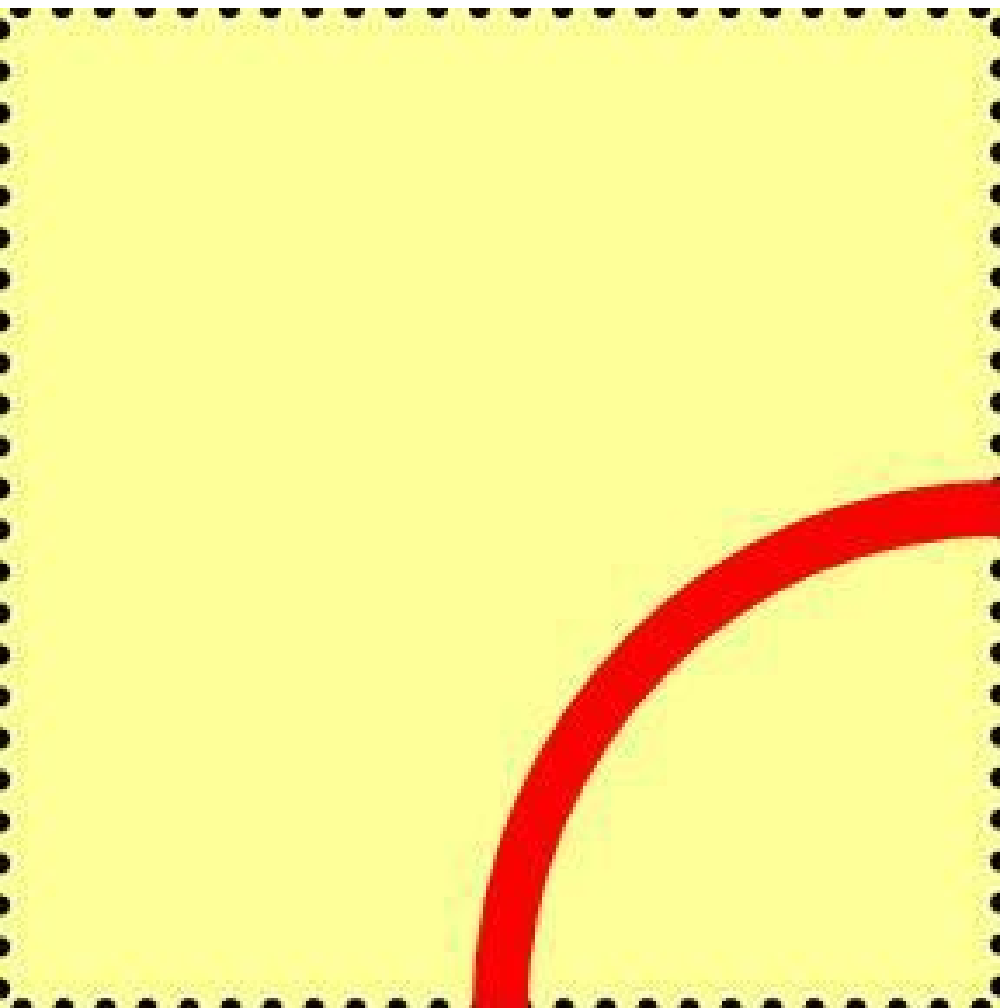} & 
\includegraphics[width=1cm]{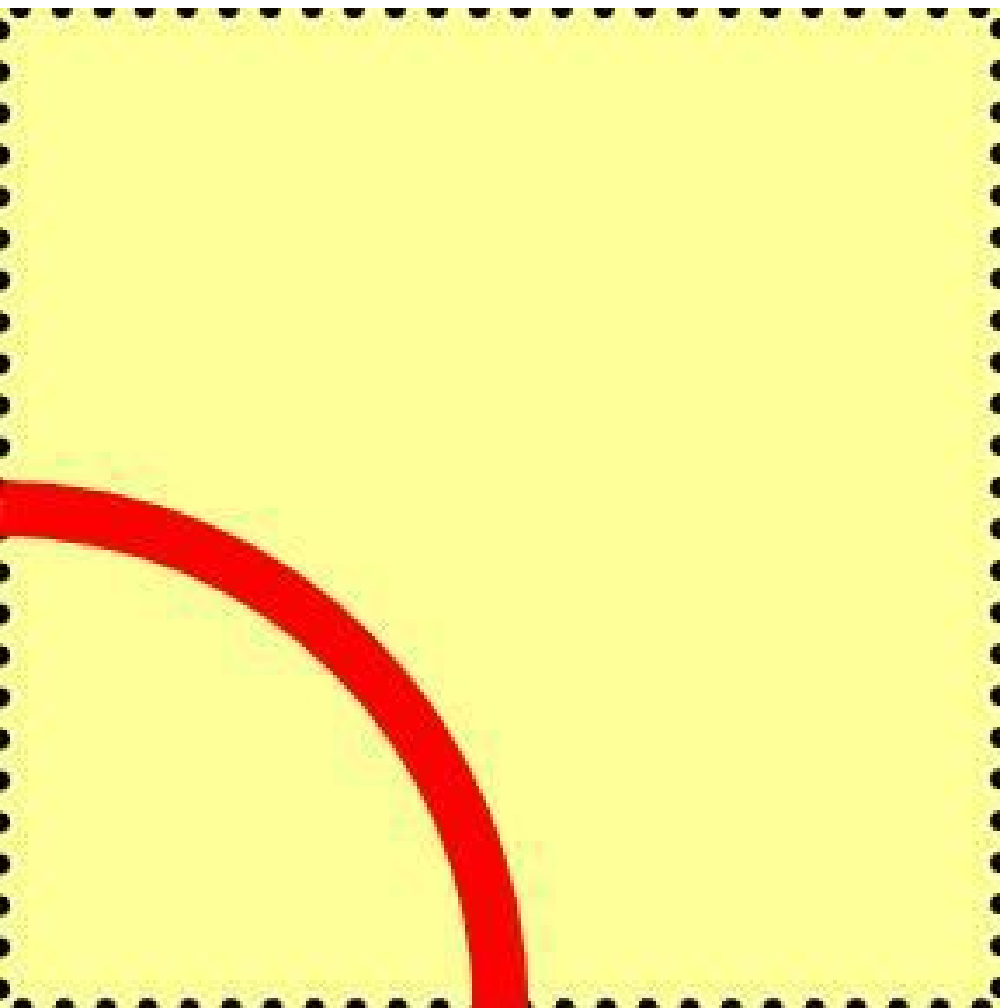} &\includegraphics[width=1cm]{ut00.EPS}  \\ 
\includegraphics[width=1cm]{ut02.EPS} & \includegraphics[width=1cm]{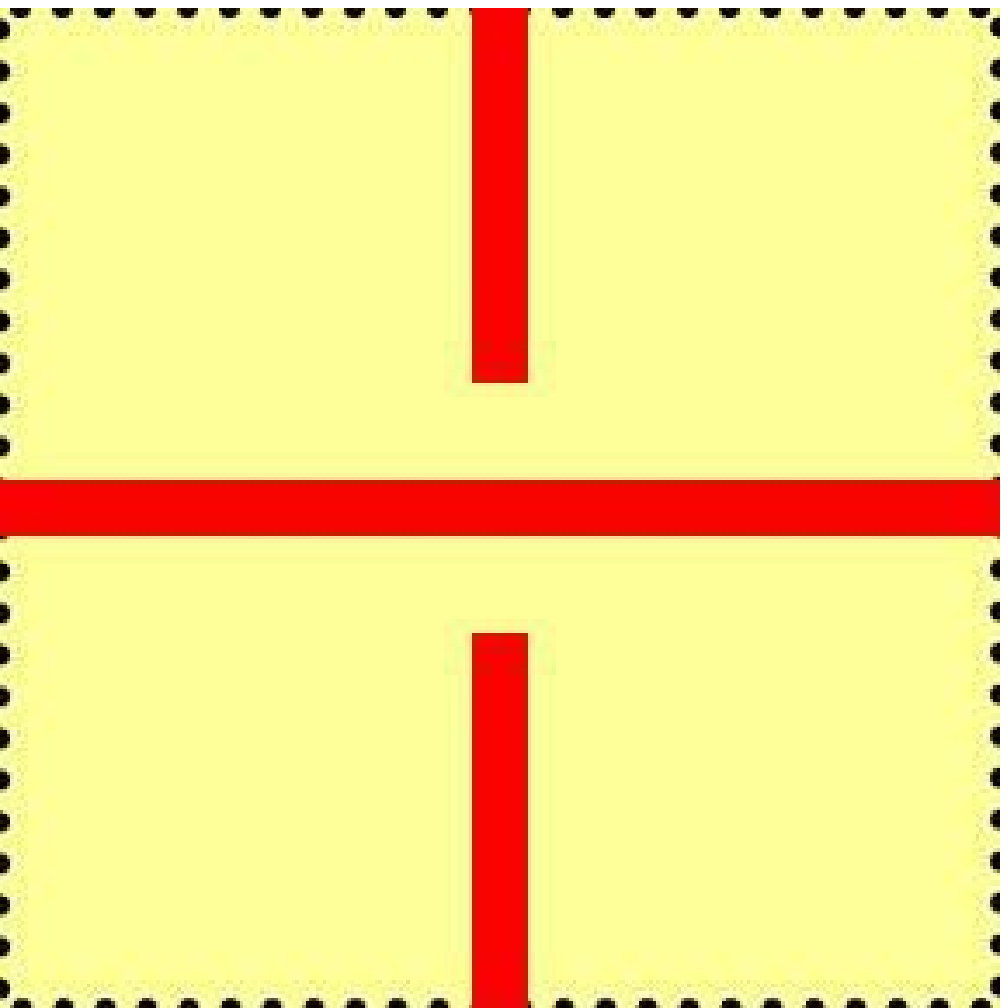} &
\includegraphics[width=1cm]{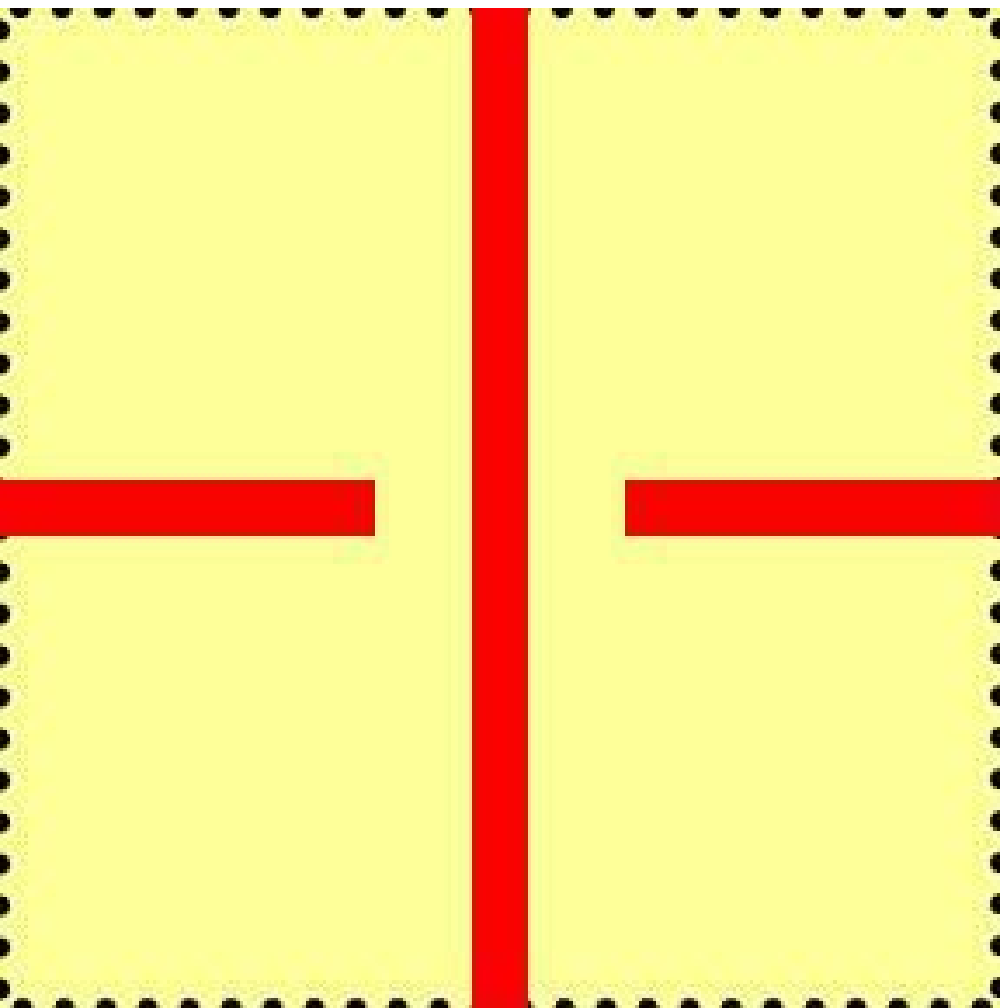} &\includegraphics[width=1cm]{ut01.EPS}  \\ 
\includegraphics[width=1cm]{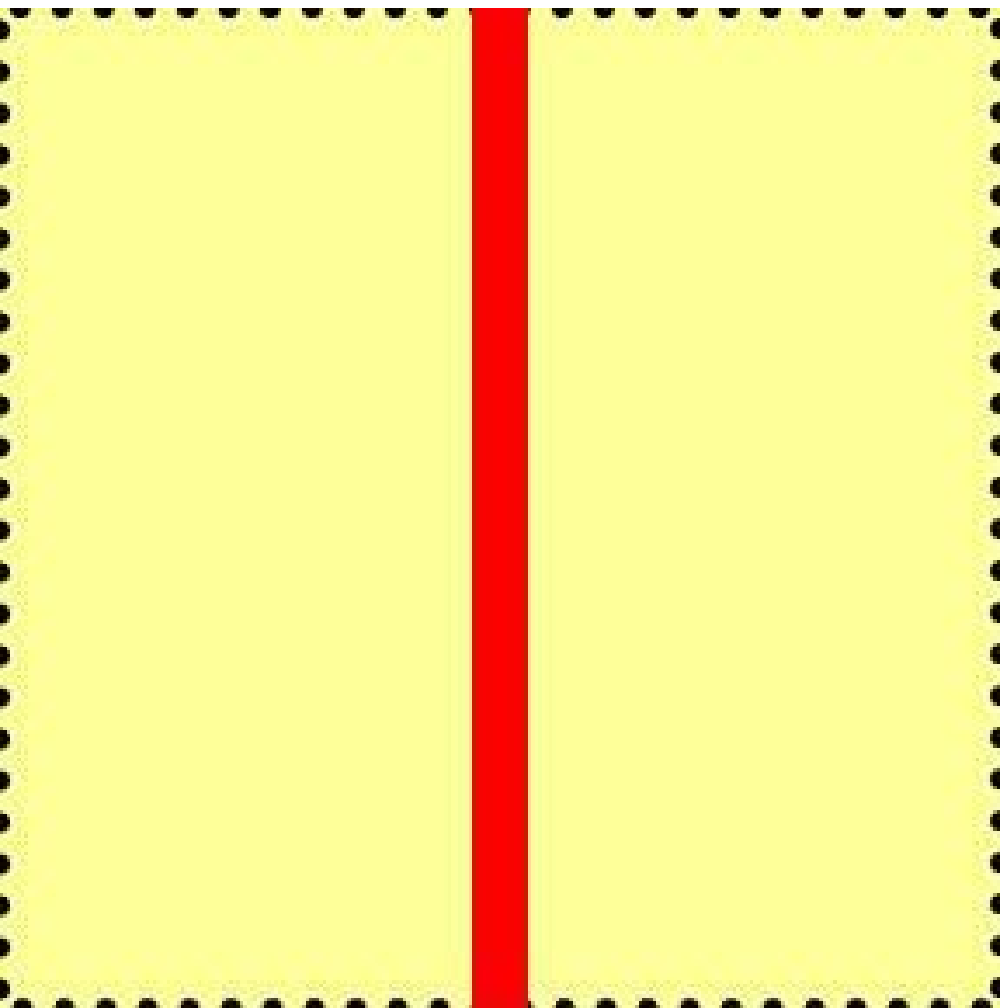} & \includegraphics[width=1cm]{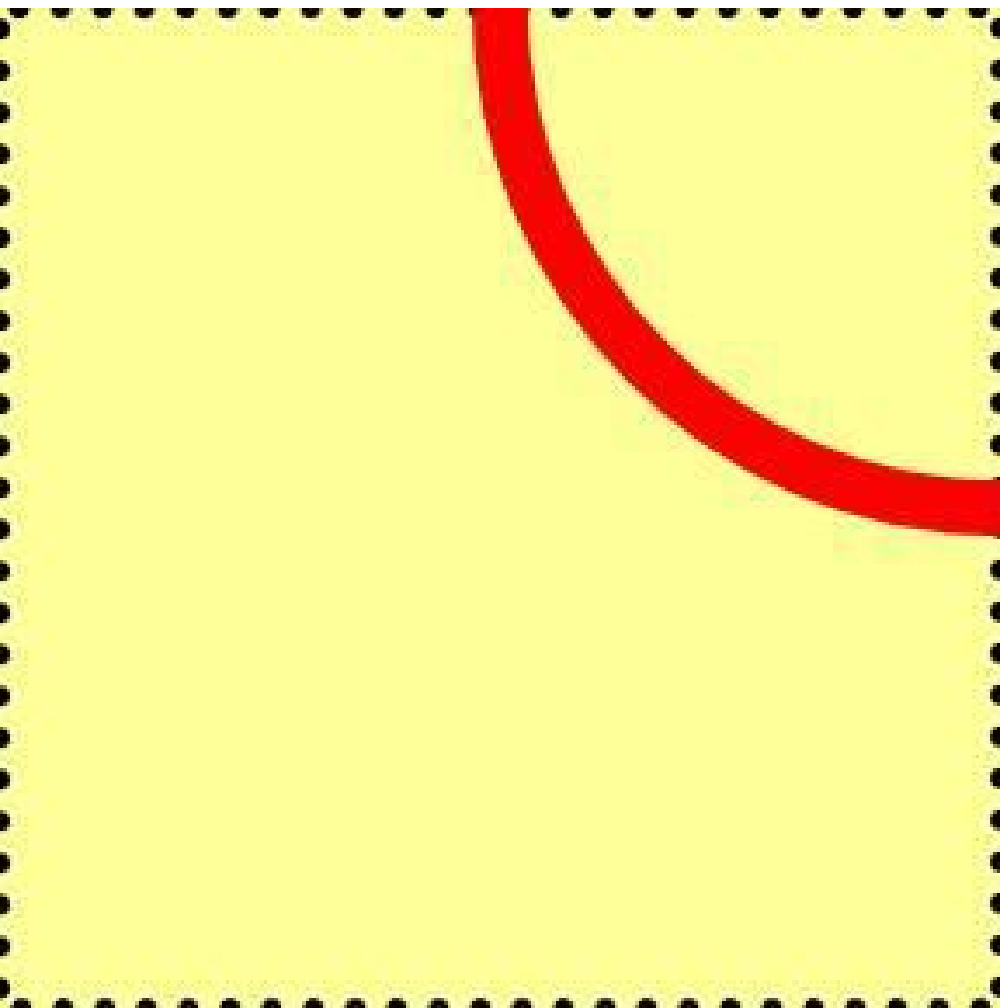} &
\includegraphics[width=1cm]{ut09.EPS} &\includegraphics[width=1cm]{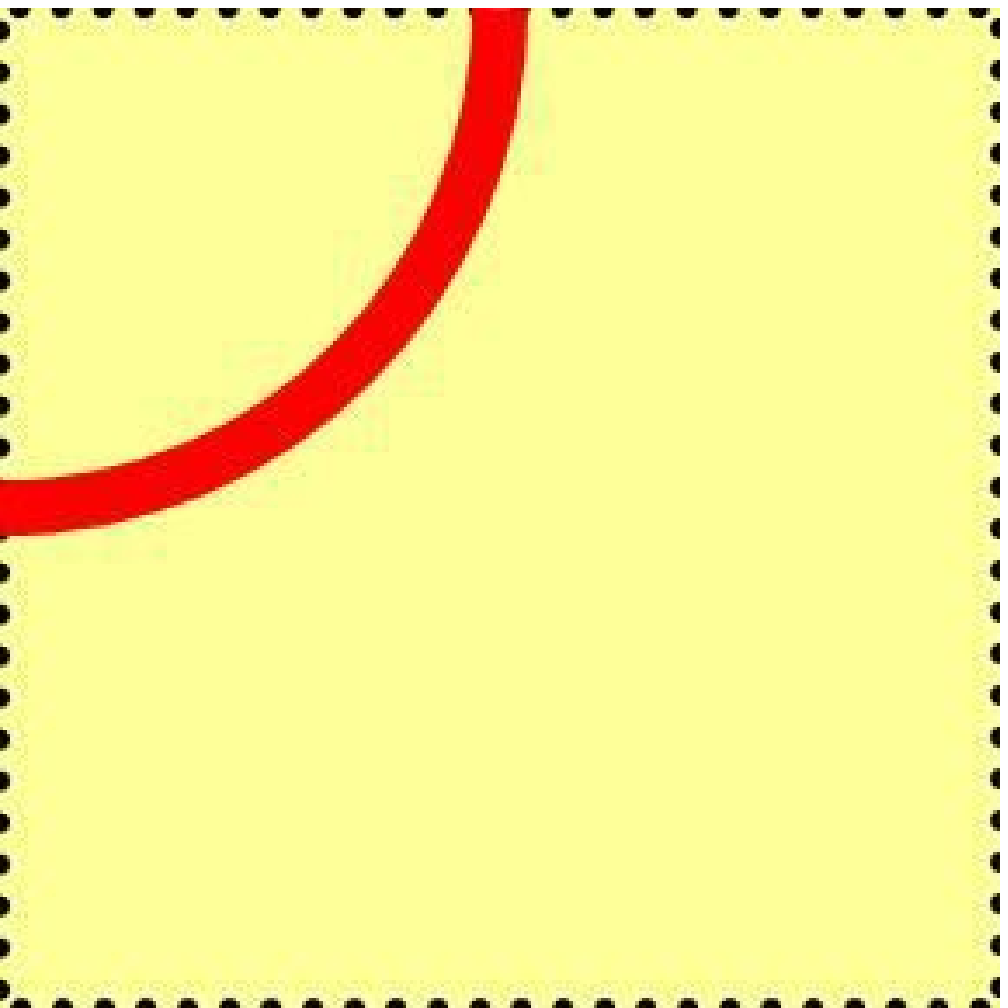}  \\ 
\includegraphics[width=1cm]{ut03.EPS} & \includegraphics[width=1cm]{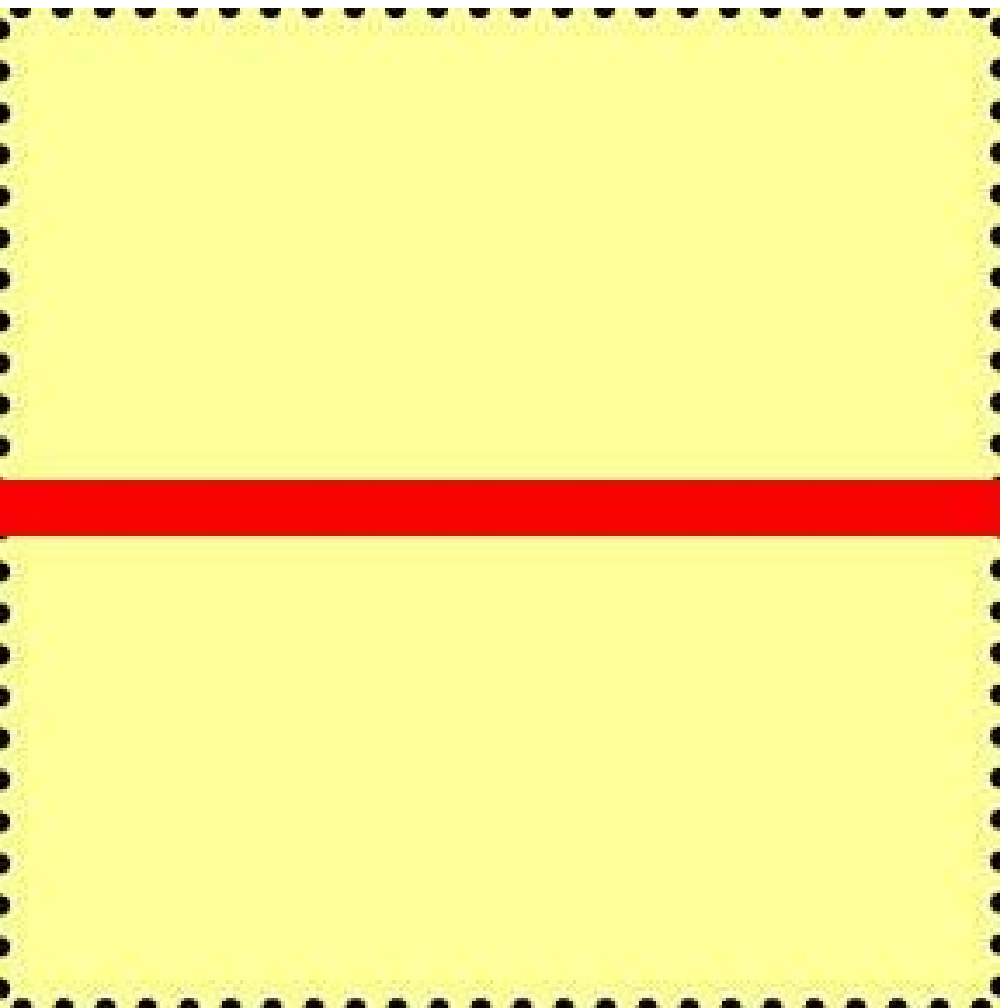} &
\includegraphics[width=1cm]{ut04.EPS} &\includegraphics[width=1cm]{ut00.EPS}   
\end{array}
$

$ 
\begin{array}{cccc}
\includegraphics[width=1cm]{ut00.EPS} & \includegraphics[width=1cm]{ut02.EPS} & 
\includegraphics[width=1cm]{ut01.EPS} &\includegraphics[width=1cm]{ut00.EPS}  \\ 
\includegraphics[width=1cm]{ut02.EPS} & \includegraphics[width=1cm]{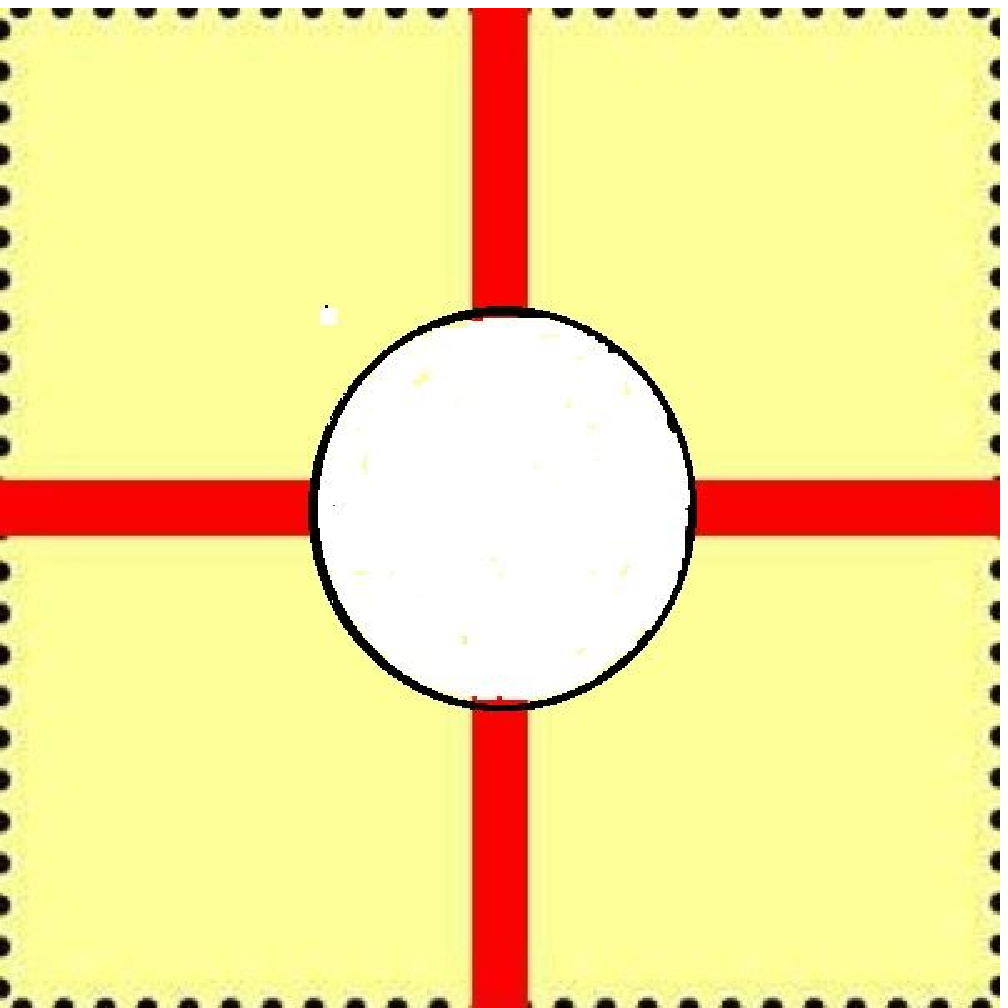} &
\includegraphics[width=1cm]{ut10.EPS} &\includegraphics[width=1cm]{ut01.EPS}  \\ 
\includegraphics[width=1cm]{ut06.EPS} & \includegraphics[width=1cm]{ut03.EPS} &
\includegraphics[width=1cm]{ut09.EPS} &\includegraphics[width=1cm]{ut04.EPS}  \\ 
\includegraphics[width=1cm]{ut03.EPS} & \includegraphics[width=1cm]{ut05.EPS} &
\includegraphics[width=1cm]{ut04.EPS} &\includegraphics[width=1cm]{ut00.EPS}   
\end{array}
$

$ 
\begin{array}{cccc}
\includegraphics[width=1cm]{ut00.EPS} & \includegraphics[width=1cm]{ut02.EPS} & 
\includegraphics[width=1cm]{ut01.EPS} &\includegraphics[width=1cm]{ut00.EPS}  \\ 
\includegraphics[width=1cm]{ut02.EPS} & \includegraphics[width=1cm]{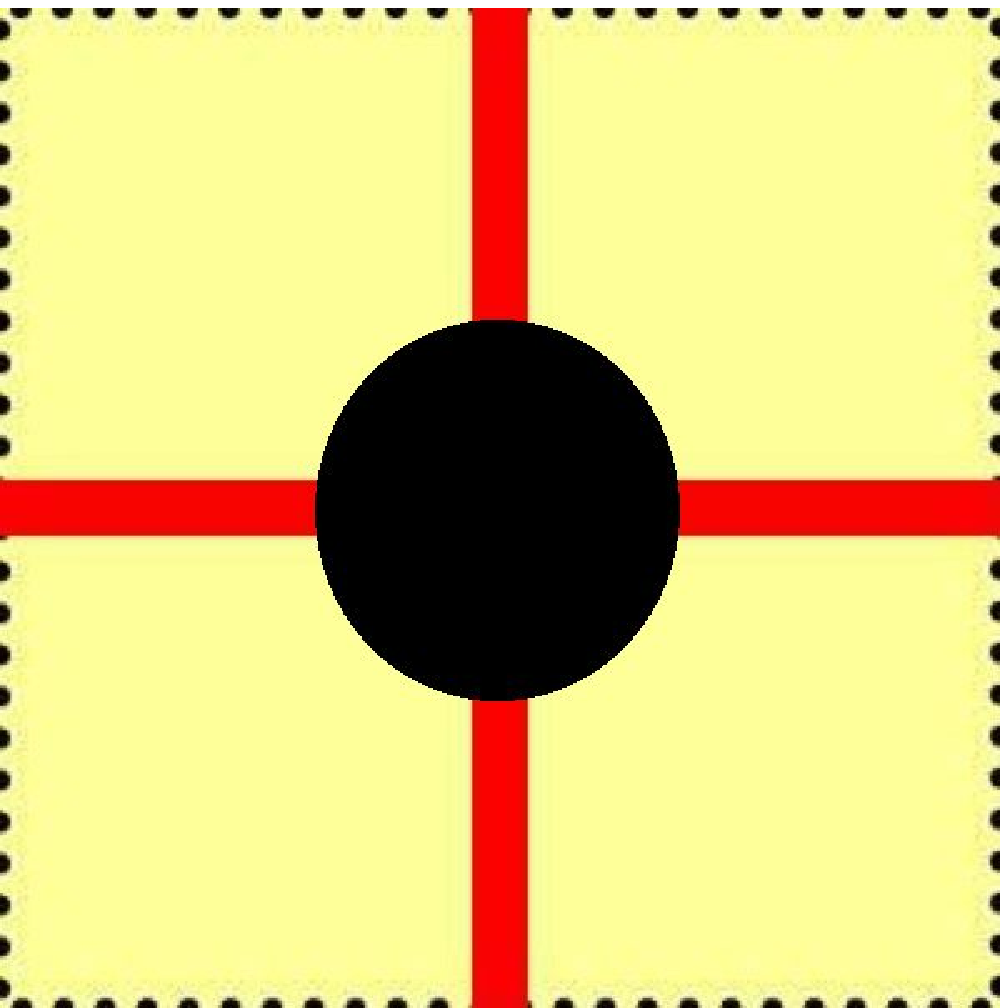} &
\includegraphics[width=1cm]{ut10.EPS} &\includegraphics[width=1cm]{ut01.EPS}  \\ 
\includegraphics[width=1cm]{ut06.EPS} & \includegraphics[width=1cm]{ut03.EPS} &
\includegraphics[width=1cm]{utvertex.EPS} & \includegraphics[width=1cm]{ut04.EPS}  \\ 
\includegraphics[width=1cm]{ut03.EPS} & \includegraphics[width=1cm]{ut05.EPS} &
\includegraphics[width=1cm]{ut04.EPS} &\includegraphics[width=1cm]{ut00.EPS}   
\end{array}
$

     \end{tabular}
     \caption{\bf Classical, Virtual and Graphical Mosaic Knots}
     \label{Figure 1}
\end{center}
\end{figure}

\begin{figure}
     \begin{center}
     \begin{tabular}{c}
     
$ 
\begin{array}{ccccccccccc}
\includegraphics[width=1cm]{ut01.EPS} & 
\includegraphics[width=1cm]{ut02.EPS} &\includegraphics[width=1cm]{ut03.EPS}  & 
\includegraphics[width=1cm]{ut04.EPS} &\includegraphics[width=1cm]{ut05.EPS} &
\includegraphics[width=1cm]{ut06.EPS}&\includegraphics[width=1cm]{ut00.EPS} &
\includegraphics[width=1cm]{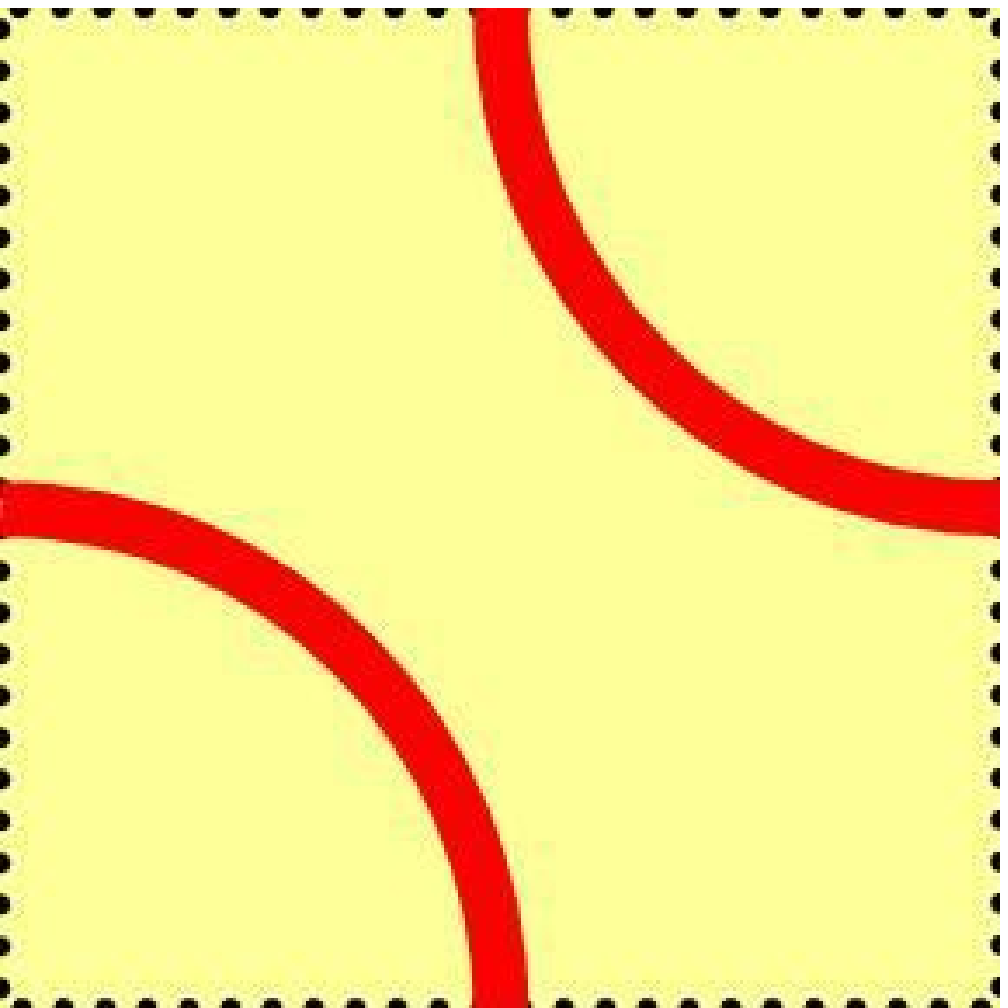}  & \includegraphics[width=1cm]{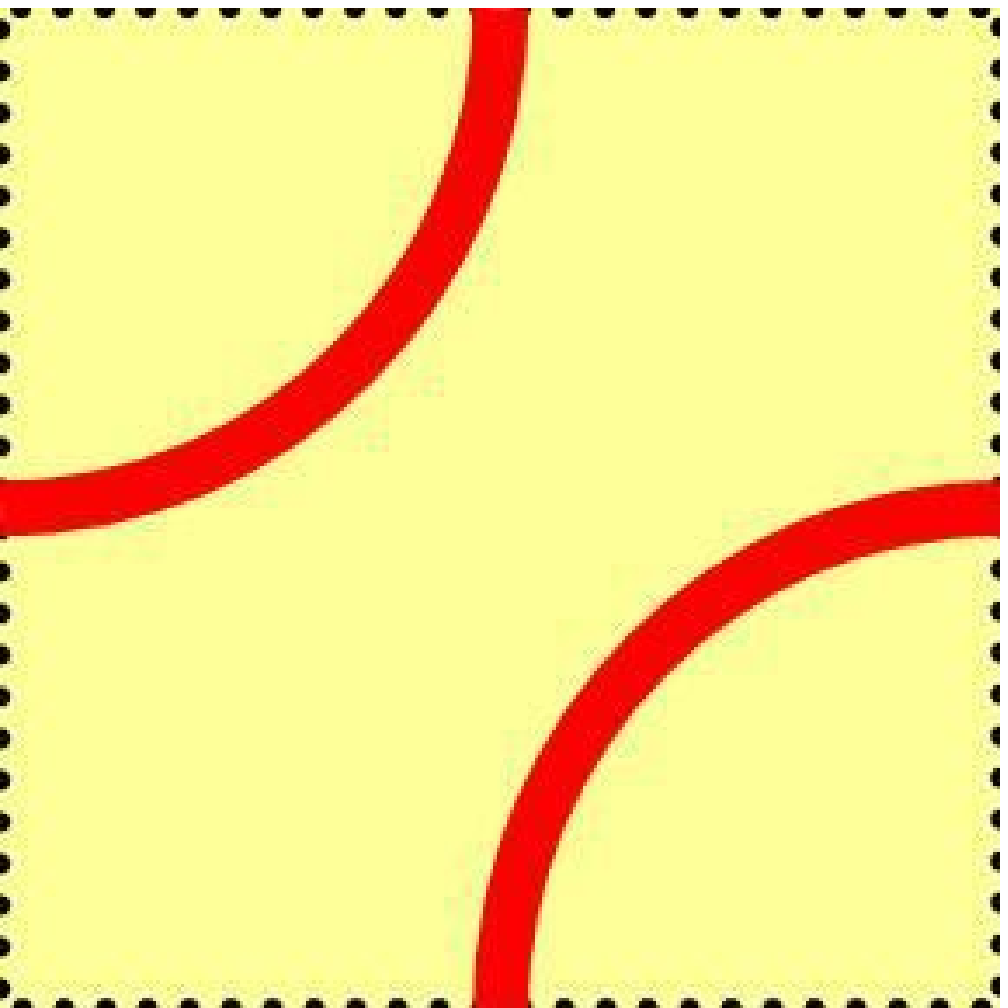} & 
\includegraphics[width=1cm]{ut09.EPS} &\includegraphics[width=1cm]{ut10.EPS}  
\end{array}
$

     \end{tabular}
     \caption{\bf Basic Tiles}
     \label{Figure}
\end{center}
\end{figure}

\section{Discussion on Quantum Knots and Entanglement}
It is worth comparing our formulation of quantum knots with the proposal of Aravind that would 
make an observable related to a link via the removal of one component of that link. This analogy is of interest in thinking about properties of entanglement. The $GHZ$ state 
$$|GHZ\rangle = (\frac{1}{\sqrt{2}})(|000\rangle + |111\rangle)$$
is entangled, but becomes unentangled upon measurement in any of its three tensor factors.
The Borommean rings are disentangled when any of the three components is removed. This suggests making a catalog of links in relation to entangled states. However, there are 
various multiplicities in this situation. See \cite{KauffTopQ} for a previous discussion about this matter. Here we
point out that there are infinitely many examples of topologically distinct three-component links with
the property that they are linked, but become unlinked upon the removal of any one component.
See Figure 3 for a depiction of the usual Borommean rings and a second example that we have called
the Companion rings. One infinite family is obtained by applying the process leading from the Borommean rings to the Companion rings recursively, or by replacing the $2 \pi$ twists in the diagram for the self-clasped components by a twists of type $2 \pi n$ for $n$ any natural number.  
It may be that this multiplicity is not the central issue in understanding relationships between topological entanglement and quantum entanglement. The key conceptual point is that topological relationships are independent of distance and quantum entanglement relationships are also independent of distance.
Thus one expects that there should be a connection between topology and quantum non-locality.
This conceptual line will be pursued elsewhere.
\bigbreak

\begin{figure}[htb]
     \begin{center}
     \begin{tabular}{c}
     \includegraphics[width=6cm]{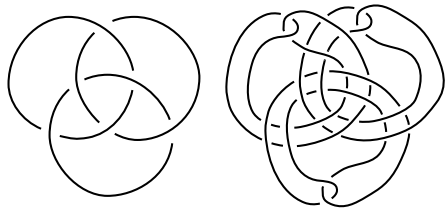}
      \end{tabular}
     \caption{\bf Borommean Rings and Companion Rings}
     \label{Figure}
\end{center}
\end{figure}

The reader will note that the notion of component removal as an observable does not occur in
our formulation of knots as quantum knots. Component removal is an operation on classical knots and 
we can consider whether there is a way to formulate it in our category of mosaic quantum knots.
It is an open problem whether this can be done in a satisfactory way. The question leads to a number of
related questions about operations on mosaic knots. Following analogies from classical knots, we can
consider operations such as: remove a component, switch a crossing, smooth a crossing. As we look
at operations that one can perform on knot and link diagrams we see that many local operations can be regarded as unitary transformations on the mosaic lattice Hilbert space. For example, switching a single 
crossing amounts to a simple permutation of basis elements in a single tensor factor of the Hilbert space
and hence is a unitary transformation. Smoothing a crossing as in 
$$\Across  \longrightarrow  \Asmooth $$
 is also a unitary transformation. Thus we have a complex situation where operations that 
change the topology of classical knots and links can be construed as unitary operations on quantum knots. The topology preserving operations that correspond to the ambient group are a special 
subset of the whole group of unitary operations.
\bigbreak

Our formulation of quantum knots does allow the superposition of diagrammatic mosaics and 
hence allows the possibility of entangled states. Such entangled quantum knots exist, and it is now 
a curious puzzle to see how topological entanglement and quantum entanglement coexist in this
category of quantum knots.

\section{General Quantization and Quantizing Classical Knots}
In this section we give a general definition of quantization, in analogy to that given in 
\cite{LomQknots2}. We then apply this definition to the quantization of classical knots that are represented by embeddings of the circle in Euclidean three-space.
\bigbreak

\noindent {\bf Definition.} Let $E$ be a collection of mathematical objects. We will call $E$ the set of 
{\em motifs} to be quantized (see \cite{LomQknots2}). Let $G$ be a group acting on the elements of 
$E$ so that each element of $G$ permutes $E.$ That is, we assume that for each $g \in G$ we have a
mapping taking $K \longrightarrow g(K)$ for each $K \in E$ such this is a 1-1 correspondence of 
$E$ with itself and so that $g(h(K)) = (gh)(K)$ where $g$ and $h$ are in $G$ and $gh$ denotes the product of these group elements in $G.$ We further assume that the identity element in $G$ acts as the identity mapping on $E.$ We then {\em quantize} the pair $(E,G)$ by forming a Hilbert space $H(E)$
with orthonormal basis consisting  in the set  $\{ |K\rangle : K \in E  \}.$  Here we take the elements of  
$H(E)$ to be finite sums of basis elements with complex coefficients and we use the usual Hermitian inner product on this space. Since the group $G$ acts on the basis by permuting it, we see that the action extends to an action of $G$ on $H(E)$ by unitary transformations. We call the new pair
$(H(E), G)$ (with this unitary action) the {\em quantization} of $(E,G).$
\bigbreak

Classical knot theory is formulated in terms of continuous embeddings of circles into the three dimensional space $R^{3}$ or the three dimensional sphere $S^{3}$ (which may be taken as the
set of vectors of unit length in Euclidean four dimensional space, or as the one-point compactification of 
$R^{3}.$ A knot is represented by an embedding $K: S^{1} \longrightarrow R^{3}.$ where $S^{1}$ denotes the circle (i.e. the set of points at unit distance from the origin in the Euclidean plane) with the 
topology inherited from the Euclidean plane. If $h:R^{3} \longrightarrow R^{3}$ is an orientation 
preserving homeomorphism of $R^{3}$, then by forming the composition $K' = h \circ K$ defined by
$h \circ K (x) = h(K(x))$ for $x \in S^{1}$, we obtain a new embedding$K'.$ We say that the two
embeddings $K$ and $K'$ are {\em equivalent}. We say that two embeddings $K$ and $K'$ represent the same knot type if there is an orientation preserving homeomorphism $h$ (as above) such that 
$K' = K\circ h.$ The set $G$ of orientation preserving homeomorphisms of $R^{3}$ forms a group under compositiion. The set of circle embeddings $$E(S^{1}) = \{K:S^{1} \longrightarrow R^{3}\}$$
is acted upon by $G$ via composition. In this way the  group $G$ acts as a group of permutations of the set $E(S^{1}).$ Note that we mean this action in the sense of group representations. We have that for
$g, h \in G,$ $$g\circ(h \circ K)) = (g\circ h)\circ K$$ for $K:S^{1} \longrightarrow R^{3},$ any embedding of a circle in $R^{3}.$ Note also that two elements $K$ and $K'$ of $E(S^{1})$ are equal if and only if they are point-wise equal as functions on the circle $S^{1}.$
\bigbreak

Let $H(E(S^{1})$ denote the Hilbert space for which the set of embeddings $E(S^{1})$ is an orthonormal basis. We take this space to be the set of finite linear combinations of its basis elements. We denote the basis elements of this Hilbert space by $|K \rangle$ where $K$ is an embedding of the circle in $R^{3}$. Using $G$ as defined above, we have $G$ applied to the basis elements of $H(E(S^{1})$ acting as a group of permutations of the basis. These permuations extend to unitary transformations on the entire Hilbert space, giving a quantization of $(E(S^{1}, G)$ in accord with the definition given in this section.
\bigbreak

\noindent {\bf Remark.} See Figure 4 for an illustration of a classical knot equivalence. Note that the
quantization of the set embeddings that represent classical knots gives a Hilbert space of uncountable 
dimension, just as there are an uncountable number of embeddings that can represent knots in three dimensional space. Thus this quantization must be contrasted with the mosiac knots where we have created a hierarchy of finite dimensional spaces and finite groups to handle quantum information for combinatorial knot theory. The quantization of classical knots that is given in this section is 
intellectually satisfying since it quantizes the full geometrical context for knot theory. This same 
context of embeddings of objects or placements of structures in three dimensional space is the place where most ideas in geometry, topology and physics are carried out. Thus we expect that this very large quantization of knots will be useful in studying knots in physical situations such as vortices in 
super-cooled helium \cite{Rasetti} or the possibility of knotted structures in gluon fields \cite{BK}.
\bigbreak 

\begin{figure}[htb]
     \begin{center}
     \begin{tabular}{c}
     \includegraphics[width=7cm]{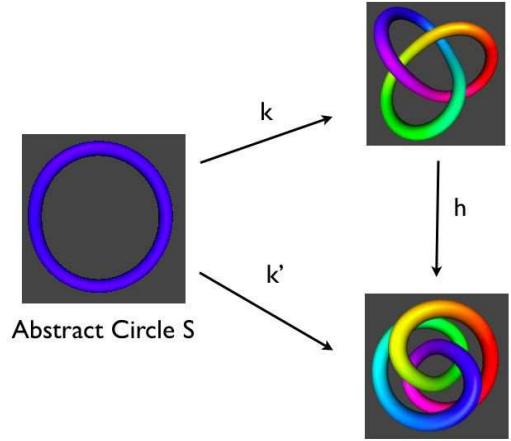}
      \end{tabular}
     \caption{\bf Classical knot equivalence via ambient homeomorphism}
     \label{Figure}
\end{center}
\end{figure}

\section{Quantum Gauss Codes}
A {\em Gauss code} is a method of recording the information inherent in a classical knot diagram.
One orients the diagram and labels the crossings in the diagram. Then, choosing a starting point one 
traverses the diagram, noting the crossing, whether one is traversing the overcrossing part of the
crossing and noting the sign of the crossing. One makes an ordered sequence of this information.
Thus the right-handed trefoil knot has the code $$o1+u2+o3+u1+o2+u3+,$$ where the crossings are labeled $1,2,3.$ Here the letters {\em o} and {\em u} stand respectively for {\em over} and 
{\em under,} and all the crossings in the trefoil knot are positive.
\bigbreak

For a fixed positive integer $N,$ let $V = V[N]$ denote the Hilbert space with basis 
$$\{ |oi+\rangle,  |oi-\rangle,  |ui+\rangle,  |ui-\rangle,  |*\rangle | i = 1,\cdots , N \}.$$
Quantum Gauss diagrams are regarded as elements of $$H = H[N,M] = V[N]^{\otimes M}.$$
These correspond to (generalized) Gauss codes that include the blank symbol $*.$
For example the code $$o1+u2+o3+u1+o2+u3+$$ corresponds to 
$$|o1+ \rangle |u2+\rangle |o3+\rangle |u1+\rangle |o2+\rangle |u3+\rangle .$$
We need to specify $N$ to allow a sufficient supply of crossings and $M$ to allow blanks to appear
when a Reidemeister move on the code removes a crossing. We now specify the Reidemeister moves.
\bigbreak

\noindent {\bf Reidemeister Moves for Quantum Gauss Codes}
\begin{enumerate}
\item The first Reidemeister move allows the replacement of 
$| oi+\rangle |ui+\rangle$ with $|*\rangle |*\rangle$ , the replacement
of $| ui+\rangle |oi+\rangle$ with $|*\rangle |*\rangle$, the replacement
of $| oi-\rangle |ui-\rangle$ with $|*\rangle |*\rangle$, and the replacement
of $| ui-\rangle |oi-\rangle$ with $|*\rangle |*\rangle$
when these factors occur successively in a tensor product. Since, in any given tensor factor,
this transformation amounts to the exchange of two basis vectors, the move is a unitary transformation
on the Hilbert space $H[N,M].$ Reidemeister moves are alllowed to go in either direction.
Thus, one can replace $|*\rangle |*\rangle$ with $| oi+\rangle |ui+\rangle.$
\item The second Reidemeister move allows the replacement of 
$| oi+\rangle |oj-\rangle X |ui +\rangle |uj - \rangle$ with 
$|*\rangle |*\rangle X |*\rangle |*\rangle$ and the replacement of 
$| oi+\rangle |oj-\rangle X |uj -\rangle |ui + \rangle$ with 
$|*\rangle |*\rangle X |*\rangle |*\rangle$
in any larger string in which this pattern appears.
Here $X$ is an arbitrary tensor product of basis elements of $V$ and $i \ne j.$
As in the case of the first move, there are variants of the second move that still hold, obtained by 
reversing all plus signs to minus signs and all minus signs to plus signs.
\item The third Reidemeister move allows the replacement of
$$|ui+ \rangle| uj+\rangle X |oi+ \rangle|uk+ \rangle Y|oj+ \rangle|ok+ \rangle$$
with
$$|uj+ \rangle|ui+ \rangle X |uk+ \rangle| oi+\rangle Y|ok+ \rangle|oj+ \rangle.$$
Here $X$ and $Y$ are arbitrary intermediate tensor factors. Note that the move takes the form of
replacing 
$$|a \rangle| b\rangle X |c \rangle |d \rangle Y| e \rangle |f \rangle$$
with
$$|b \rangle| a\rangle X |d \rangle |c \rangle Y| f \rangle | e\rangle$$for special choices of 
$a,b,c,d,e,f.$ There are a number of variations of this move. We omit  the remaining variations,
trusting them to the reader familiar with the classical knot theoretic Reidemeister three move.
\item We include two more quantum moves that take care of the appearance of the blanks.
The first of these is the interchange of $|a \rangle |* \rangle$ with $|* \rangle |a \rangle$
for any basis vector $|a \rangle$ of $V.$ The second is the cyclic permuation 
replacing $|a_{1}\rangle |a_{2}\rangle\cdots |a_{M}\rangle$ with 
 $|a_{M}\rangle |a_{1}\rangle\cdots |a_{M-1}\rangle.$ where the $|a_{k}\rangle$ are any basis vectors
 for $V.$
\end{enumerate}

\noindent {\bf Remark.} In Figure 5 we illustrate diagrams for the Reidemeister moves. In the first
part of the figure we show the diagrams for the moves without choice of orientation. In the second
part of the figure we show diagrams with orientation and labels that correspond to the notations 
used above. In the case of the third Reidemeister move we have chosen the orientations and the
order of traverse of the diagram to correspond to the special case noted above. The reader should 
now be able to see how to enumberate all possible cases of the third move.
\bigbreak

\begin{figure}[htb]
     \begin{center}
     \begin{tabular}{c}
     \includegraphics[width=6cm]{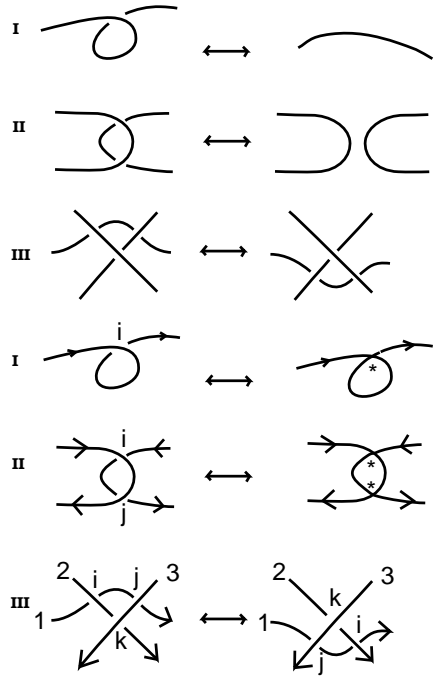}
      \end{tabular}
     \caption{\bf Reidemeister Moves}
     \label{Figure}
\end{center}
\end{figure}

\noindent {\bf Remark.} Along with the Reidemeister moves, we also want to allow permutations
of the indices that mark the sites in the Gauss code. Such permutations induce permutations of the 
basis elements in $V[N]$ and hence induce unitary transformations on $H[N,M]$ that preserve the 
topological type of the knots and links in this combinatorial level.
\bigbreak

\noindent {\bf Remark.} Some caveats are important in using these moves. When a move is applied in the direction that exchanges blank elements with non-blank elements, the  indices $i$ that appear
(as in $|*\rangle |*\rangle \longrightarrow | oi+\rangle |ui+\rangle$) must be new indices that are not
already in the tensor product. This means that there is a limitation to the performance of this type of move
if we take $N$ to be finite. If we wish to use infinite dimensional spaces, then there is no limitation.
In the case of infinite dimensional spaces for the indices the basic space will be denoted by
$V[\infty].$ Since in doing a move in a direction such as  $|*\rangle |*\rangle \longrightarrow | oi+\rangle |ui+\rangle$ one needs the existence of appropriate blanks in the tensor product, some moves that a
classical knot theorist might want are not neccesarily available. This is the limitation of the number of
the $M$ tensor factors in the Hilbert space $H[N,M].$ Again, we can eliminate this restriction by letting
$M$ be infinite. The individual vectors will then be of the form 
$$|a_{1}\rangle |a_{2} \rangle \cdots |a_{k} \rangle |* \rangle |* \rangle |*\rangle \cdots $$
where all factors are blank beyond some finite but unlimited number of factors. In this case one can apply the cyclic permutation to the first $k$ (as above) factors. 
\bigbreak

With these remarks, the space of quantum Gauss
codes has a an ambient  group of generalized Reidemeister moves acting unitarily upon it and we can do quantum knot theory
in this category. The Theorems we have proved in the previous section on observables and invariants 
of mosaic quantum knots apply mutatis mutandis to the quantum Gauss codes. The codes provide 
a combinatorial way to handle problems in knot theory and they are very close to both computational
aspects and to the theory of invariants. One advantage in using the Gauss code formalism for quantum
knots is that one can depict them with the usual knot diagrams, however the extra appearance of blanks
in the quantum codes must be made to correspond to flattened crossings in the diagrams as is illustrated 
in Figure 5. Much more remains to be done in this domain.
\bigbreak

\section{Quantum Graphs}
In \cite{LomQknots1}, in section 28, we consider the concept of a quantum graph where the graph is a lattice  graph in either a planar lattice (the context for mosaic quantum knots) or a three  dimensional lattice
(the context for quantum knots in a three dimensional form). In this section we point out a generalization 
that is implicit in our original definition, allowing a quantization of arbitrary abstract graphs. 
\bigbreak

Let $G$ be a finite simple directed graph in the sense of graph theory. This means that $G$ consists in a finite set of {\em nodes (vertices)} , denoted by $V(G)$ and a finite set of {\em edges} denoted by $E(G)$ such that every edge is associated with two distinct nodes of $G$ in the form of an orderer pair
$(a,b)$ where $a$ and $b$ are distinct elements of the set of vertices $V(G).$ We will denote an edge by its corresponding ordered pair. Note that given two distinct nodes $a$ and $b$ there are up to two 
edges associated with them: $(a,b)$ and $(b,a).$ A graph is completely specified by giving the 
vertex set $V(G)$ and a set of ordered pairs of distinct nodes, specifying $E(G).$
Two graphs are {\em isomorphic} if there is a $1-1$ correspondence between their sets of nodes that induces a  $1-1$ correspondence between their sets of edges. 
\bigbreak

Let a graph $G$ be specified by $V(G) = \{1,2, \cdots , n\}$ for some positive integer $n,$ with $E(G)$  a subset of the set of ordered pairs $(i,j)$ where $1 \le i \ne j \le n.$ Define a complex vector space 
$H_{n}$ of dimension $n(n-1)$ with basis the set of all kets of the form $| (i,j) \rangle$ where 
$1 \le i \ne j \le n.$  Let $d(G)$ denote the cardinality of $E(G)$ and define 
$|G\rangle \in H_{n}^{\otimes d(G)}$ by the formula $$|G\rangle = \bigotimes_{(a,b) \in E(G)} |(a,b)\rangle$$ where the order of this tensor product is the lexicographic order of the set of ordered pairs in $E(G).$
This formula defines a state vector in $H_{n}^{\otimes d(G)}$ corresponding to the graph $G.$ In this way we can formulate {\em quantum graphs} just as we have formulated quantum knots. Graph isomorphisms are induced by appropriate permutations of the set  $V(G) = \{1,2, \cdots , n\},$ and induce
unitary transformations on the space $H_{n}^{\otimes d(G)}.$ We will stop here at this point of definition,
and continue the analysis of quantum graphs in a separate paper.
\bigbreak

\section{Quantizing Words in Groups}
In this section the letter $G$ will stand for a finitely presented and finitely related group.
If $G$ has generators $\{ x_{1}, \cdots, x_{n} \}$ then we can consider words in the group $G$
as products of these generators including an identity symbol $*.$ Then we can represent a
word in the group by a corresponding tensor product of kets. For example
$x_{1}x_{2}^{-1} **x_{3}$ will correspond to the ket 
$$|x_{1}\rangle |x_{2}^{-1}\rangle | *\rangle |*\rangle |x_{3} \rangle .$$
In this way, words in the group presentation correspond to elements of a complex tensor product of a 
space spanned by the kets $|x_{i}\rangle$, $|x_{i}^{-1}\rangle$ and $|*\rangle.$ The same considerations that we have made in the previous constructions allow us to define unitary transformations corresponding to applications of the group laws and the relations in the group.
Finite dimensional tensor products hold only words up to a given length, but we can also work with 
the full set of words in the group by using infinite tensor products of the base space.
\bigbreak

Note that essentially the same construction can be applied to any algebraic system with a single (semi)  binary operation. This includes the possibility of quantizing sequences of composable morphisms in 
an arbitrary category. Note  also that in making a construction of this kind, we actually quantize all the words in the group.Such quantizations have at the present time little to do with the subject of so-called quantum groups. In the case of the Artin braid group this quantization gives ``quantum braids", a structure precisely analogous to our quantum knots, and treated separately in another paper in these proceedings \cite{LomQknots2} by 
S. J. Lomonaco and the present author.
\bigbreak

\section{Background on the Bracket Polynomial and Jones Polynomial}
In the next sections we show how a different approach to forming Hilbert spaces corresponding
to combinatorial data leads to a quantum algorithm for the Jones polynomial and to relations
with Khovanov homology. This work of the first author appears in \cite{KhoJones}. In this mode, we first explain the state summation for the bracket polynomial
model for the Jones polynomial, and then we make a Hilbert space that has basis the set of enhanced 
states for the bracket (to be defined below). There is a natural unitary transformation on this space that
encodes the bracket polynomial and gives rise to the corresponding quantum algorithm.
\bigbreak

The bracket polynomial \cite{KaB} model for the Jones polynomial \cite{JO,JO1,JO2,Witten} is usually described by the expansion
$$\langle \Across \rangle=A \langle \Asmooth \rangle + A^{-1}\langle \Bsmooth \rangle$$
Here the small diagrams indicate parts of otherwise identical larger knot or link diagrams. The two types of smoothing (local diagram with no crossing) in
this formula are said to be of type $A$ ($A$ above) and type $B$ ($A^{-1}$ above).

$$\langle \bigcirc \rangle = -A^{2} -A^{-2}$$
$$\langle K \, \bigcirc \rangle=(-A^{2} -A^{-2}) \langle K \rangle $$
$$\langle \Rcurl \rangle=(-A^{3}) \langle \Arc \rangle $$
$$\langle \Lcurl \rangle=(-A^{-3}) \langle \Arc \rangle $$
One uses these equations to normalize the invariant and make a model of the Jones polynomial.
In the normalized version we define $$f_{K}(A) = (-A^{3})^{-wr(K)} \langle K \rangle / \langle \bigcirc \rangle $$
where the writhe $wr(K)$ is the sum of the oriented crossing signs for a choice of orientation of the link $K.$ Since we shall not use oriented links
in this paper, we refer the reader to \cite{KaB} for the details about the writhe. One then has that $f_{K}(A)$ is invariant under the Reidemeister moves
(again see \cite{KaB}) and the original Jones polynonmial $V_{K}(t)$ is given by the formula $$V_{K}(t) = f_{K}(t^{-1/4}).$$ The Jones polynomial
has been of great interest since its discovery in 1983 due to its relationships with statistical mechanics, due to its ability to often detect the
difference between a knot and its mirror image and due to the many open problems and relationships of this invariant with other aspects of low
dimensional topology.
\bigbreak

\noindent {\bf The State Summation.} In order to obtain a closed formula for the bracket, we now describe it as a state summation.
Let $K$ be any unoriented link diagram. Define a {\em state}, $S$, of $K$  to be the collection of planar loops resulting from  a choice of
smoothing for each  crossing of $K.$ There are two choices ($A$ and $B$) for smoothing a given  crossing, and
thus there are $2^{c(K)}$ states of a diagram with $c(K)$ crossings.
In a state we label each smoothing with $A$ or $A^{-1}$ according to the convention
indicated by the expansion formula for the bracket. These labels are the  {\em vertex weights} of the state.
There are two evaluations related to a state. The first is the product of the vertex weights,
denoted $\langle K|S \rangle .$
The second is the number of loops in the state $S$, denoted  $||S||.$

\noindent Define the {\em state summation}, $\langle K \rangle $, by the formula

$$\langle K \rangle  \, = \sum_{S} <K|S> \delta^{||S||}$$
where $\delta = -A^{2} - A^{-2}.$
This is the state expansion of the bracket. It is possible to rewrite this expansion in other ways. For our purposes in
this paper it is more convenient to think of the loop evaluation as a sum of {\it two} loop evaluations, one giving $-A^{2}$ and one giving
$-A^{-2}.$ This can be accomplished by letting each state curve carry an extra label of $+1$ or $-1.$ We describe these {\it enhanced states}
below.
\bigbreak

\noindent {\bf Changing Variables.} Letting $c(K)$ denote the number of crossings in the diagram $K,$ if we replace $\langle K
\rangle$ by
$A^{-c(K)} \langle K \rangle,$ and then replace $A^2$ by $-q^{-1},$ the bracket is then rewritten in the
following form:
$$\langle \Across \rangle=\langle \Asmooth \rangle-q\langle \Bsmooth \rangle $$
with $\langle \bigcirc\rangle=(q+q^{-1})$.
It is useful to use this form of the bracket state sum
for the sake of the grading in the Khovanov homology (to be described below). We shall
continue to refer to the smoothings labeled $q$ (or $A^{-1}$ in the
original bracket formulation) as {\it $B$-smoothings}.
\bigbreak

\noindent {\bf Using Enhanced States.}
We now use the convention of {\it enhanced
states} where an enhanced state has a label of $1$ or $-1$ on each of
its component loops. We then regard the value of the loop $q + q^{-1}$ as
the sum of the value of a circle labeled with a $1$ (the value is
$q$) added to the value of a circle labeled with an $-1$ (the value
is $q^{-1}).$ We could have chosen the less neutral labels of $+1$ and $X$ so that
$$q^{+1} \Longleftrightarrow +1 \Longleftrightarrow 1$$
and
$$q^{-1} \Longleftrightarrow -1 \Longleftrightarrow X,$$
since an algebra involving $1$ and $X$ naturally appears later in relation to Khovanov homology. It does no harm to take this form of labeling from the
beginning. The use of enhanced states for formulating Khovanov homology was pointed out by Oleg Viro in
\cite{Viro}.
\bigbreak

Consider the form of the expansion of this version of the
bracket polynonmial in enhanced states. We have the formula as a sum over enhanced states $s:$
$$\langle K \rangle = \sum_{s} (-1)^{i(s)} q^{j(s)} $$
where $i(s)$ is the number of $B$-type smoothings in $s$ and $j(s) = i(s) + \lambda(s)$, with $\lambda(s)$ the number of loops  labeled $1$ minus
the number of loops labeled $-1$ in the enhanced state $s.$
\bigbreak

One advantage of the expression of the bracket polynomial via enhanced states is that it is now a sum of monomials. We shall make use of this property
throughout the rest of the paper.
\bigbreak

\section{Quantum Statistics and the Jones Polynomial}
We now use the enhanced state summation for the bracket polynomial with variable $q$ to give a quantum formulation of the state sum.
Let $q$ be on the unit circle in the complex plane. (This is equivalent to letting the original bracket variable $A$ be on the unit
circle and equivalent to letting the Jones polynmial variable $t$ be on the unit circle.) Let ${\mathcal C}(K)$ denote the complex vector space
with orthonormal basis $ \{ |s\rangle$ \} where $s$ runs over the enhanced states of the diagram $K.$ The vector space ${\mathcal C}(K)$
is the (finite dimensional) Hilbert space for our quantum formulation of the Jones polynomial.
While it is customary for a Hilbert space to be written with the letter
$H,$ we do not follow that convention here, due to the fact that we shall soon regard ${\mathcal C}(K)$ as a chain complex and take its
homology. One can hardly avoid using  $\mathcal{H}$ for homology.
\bigbreak

\noindent With $q$ on the unit circle, we define a unitary transformation $$U: {\mathcal C}(K) \longrightarrow {\mathcal C}(K)$$ by
the formula
$$U |s \rangle = (-1)^{i(s)} q^{j(s)} |s \rangle$$ for each enhanced state $s.$ Here $i(s)$ and $j(s)$ are as defined in the previous
section of this paper.
\bigbreak

\noindent Let $$| \psi \rangle = \sum_{s} |s \rangle.$$ The state vector  $ | \psi \rangle$ is the sum over the basis states of our Hilbert space ${\mathcal
C}(K).$ For convenience, we do not normalize $|\psi \rangle$ to length one in the Hilbert space ${\mathcal C}(K).$
We then have the
\bigbreak

\noindent {\bf Lemma.} The evaluation of the bracket polynomial is given by the following formula
$$\langle K \rangle = \langle \psi | U | \psi \rangle .$$
\bigbreak

\noindent {\bf Proof.} $$\langle \psi | U | \psi \rangle = \sum_{s'} \sum _{s} \langle s'| (-1)^{i(s)}q^{j(s)} |s \rangle = \sum_{s'} \sum
_{s}(-1)^{i(s)}q^{j(s)} \langle s'| s \rangle $$
$$ =  \sum_{s}(-1)^{i(s)}q^{j(s)} = \langle K \rangle,$$ since $$\langle s'| s \rangle = \delta(s,s')$$
where $\delta(s,s')$ is the Kronecker delta, equal to $1$ when $s = s'$ and equal to $0$ otherwise. //
\bigbreak

Thus the bracket polyomial evaluation is a quantum amplitude for the measurement of the state
$U | \psi \rangle$ in the $\langle \psi |$ direction. Since $\langle \psi | U | \psi \rangle$ can be regarded as a diagonal element of the
transformation $U$ with respect to a basis containing $|\psi \rangle,$ this formula can be taken as the foundation for a quantum algorithm that computes
the bracket of $K$ via the Hadamard test. See  \cite{KhoJones} for a discussion of how the present description interfaces with the Hadamard test. The reader can examine \cite{Ah1,Ah2,QCJP,Three,Group1} for more information.
\bigbreak

It is useful to formalize the bracket evaluation as a quantum amplitude. This is a direct way to give a physical interpretation of the
bracket state sum and the Jones polynomial. Just how this process can be implemented physically depends upon the interpretation of the Hilbert
space ${\mathcal C}(K).$ It is common practice in theorizing about quantum computing and quantum information to define a Hilbert space in terms
of some mathematically convenient basis (such as the enhanced states of the knot or link diagram $K$) and leave open the possibility of a realization
of the space and the quantum evolution operators that have been defined upon it. In
principle any finite dimensional unitary operator can be realized by some physical system. In practice, this is the problem of constructing quantum
computers and quantum information devices. It is not so easy to construct what can be done in principle, and the quantum states that are produced
may be all too short-lived to produce reliable computation. Nevertheless, one has the freedom to create spaces and operators on the mathematical level
and to conceptualize these in a quantum mechanical framework. The resulting structures may be realized in nature and in present or future
technology. In the case of our Hilbert space associated with the bracket state sum and its corresponding unitary transformation $U,$ there is rich extra
structure related to Khovanov homology that we discuss in the next section. One hopes that in a (future) realization of these spaces and operators,
the Khovanov homology will play a key role in quantum information related to the knot or link $K.$
\bigbreak

\section{Khovanov Homology and a Quantum Model for the Jones Polynomial}
In this section we outline how the Khovanov homology is related with our quantum model. This can be done essentially axiomatically, without
giving the details of the Khovanov construction. See \cite{KhoJones} for more details. The outline is as follows:
\begin{enumerate}
\item There is a boundary operator $\partial$ defined on the Hilbert space of enhanced states of a link diagram $K$
$$\partial:{\mathcal C}(K) \longrightarrow {\mathcal C}(K)$$ such that $\partial \partial = 0$ and so that if
${\mathcal C}^{i,j} = {\mathcal C}^{i,j}(K)$ denotes the subspace of ${\mathcal C}(K)$ spanned by enhanced states $| s \rangle$ with $i = i(s)$ and $j = j(s),$ then
$$\partial : {\mathcal C}^{ij} \longrightarrow {\mathcal C}^{i+1,j}.$$ That is, we have the formulas $$i(\partial |s \rangle) = i(| s \rangle) + 1$$ and
$$j(\partial | s \rangle ) = j(| s \rangle )$$ for each enhanced state $s.$ Construction of the 
boundary operator can be found in \cite{KhoJones}.

\item {\bf Lemma. } By defining $U:{\mathcal C}(K) \longrightarrow {\mathcal C}(K)$ as in the previous section, via $$U|s \rangle = (-1)^{i(s)} q^{j(s)} |s
\rangle,$$ we have the following basic relationship between $U$ and the boundary operator $\partial:$ $$ U \partial + \partial  U = 0.$$
\smallbreak

\noindent {\bf Proof. } This follows at once from the definition of $U$ and the fact that $\partial$ preserves $j$ and increases $i$ to $i+1.$ //

\item  From this Lemma
we conclude that the operator $U$ acts on the homology of ${\mathcal C}(K).$ We can regard $H({\mathcal C}(K)) = Ker(\partial)/Image(\partial)$ as a new Hilbert
space on which the unitary  operator $U$ acts. In this way, the Khovanov homology and its relationship with the Jones polynomial has a natural quantum
context.

\item For a fixed value of $j$, $${\mathcal C}^{\bullet,j} = \oplus_{i} {\mathcal C}^{i,j}$$ is a subcomplex of ${\mathcal C}(K)$
with the boundary operator $\partial.$ Consequently, we can speak of the homology $H({\mathcal C}^{\bullet,j}).$ Note that the dimension of ${\mathcal C}^{ij}$ is
equal to the number of enhanced states $|s \rangle$ with $i = i(s)$ and $j = j(s).$ Consequently, we have
$$\langle K \rangle = \sum_{s} q^{j(s)}(-1)^{i(s)} = \sum_{j} q^{j} \sum_{i}(-1)^{i} dim({\mathcal C}^{ij})$$
$$= \sum_{j} q^{j} \chi({\mathcal C}^{\bullet,j}) = \sum_{j}q^{j} \chi(H({\mathcal C}^{\bullet,j})).$$ Here we use the definition of the {\it Euler characteristic of
a chain complex} $$\chi({\mathcal C}^{\bullet,j}) = \sum_{i}(-1)^{i} dim({\mathcal C}^{ij})$$ and the fact that the Euler characteristic of the complex is equal to
the Euler characteristic of its homology. The quantum amplitude associated with the operator $U$ is given
in terms of the Euler characteristics of the Khovanov homology of the link $K.$ $$\langle K \rangle = \langle \psi |U |\psi \rangle = \sum_{j}q^{j}
\chi(H({\mathcal C}^{\bullet,j}(K))).$$
\end{enumerate}

Our reformulation of the bracket polynomial in terms of the unitary operator $U$ leads to a new viewpoint on the Khovanov homology
as a representation space for the action of $U.$ The bracket polynomial is then a quantum amplitude that expresses the Euler characteristics of the
homology associated with this action. The decomposition of the chain complex into the parts ${\mathcal C}^{i,j}(K)$ corresponds to the eigenspace
decomposition of the operator $U.$ The reader will note that in this case the operator $U$ is already diagonal in the basis of enhanced states for the chain
complex ${\mathcal C}(K).$  We regard this reformulation as a guide to further questions about the relationship of the Khovanov homology with
quantum information associated with the link $K.$
\bigbreak

The internal combinatorial structure of the set of enhanced states for the bracket summation leads to the Khovanov
homology theory, whose graded Euler characteristic yields the bracket state sum. Thus we have a quantum statistical interpretation of the Euler
characteristics of the Khovanov homology theory, and a conceptual puzzle about the nature of this relationship with the Hilbert space of that quantum
theory. It is that relationship that is the subject of this paper. The unusual point about the Hilbert space is that each of its basis elements has a
specific combinatorial structure that is related to the topology of the knot $K.$ Thus this Hilbert space is, from the point of view of its basis elements, a
form of ``taking apart" of the topological structure of the knot that we are interested in studying.
\bigbreak

\noindent {\bf Homological structure of the unitary transformation.}
We now prove a general result about the structure of a chain complex that is also a finite dimensional Hilbert space.
Let ${\mathcal C}$ be a chain complex over the complex numbers with boundary operator $$\partial: {\mathcal C}^{i} \longrightarrow {\mathcal C}^{i+1},$$
with ${\mathcal C}$ denoting the direct sum of all the ${\mathcal C}^{i}$, $i = 0,1,2,\cdots n$ (for some $n$). Let
$$U:{\mathcal C} \longrightarrow {\mathcal C}$$ be a unitary operator that satisfies the equation $U \partial + \partial U = 0.$ We do not assume a second
grading $j$ as occurs in the Khovanov homology. However, since $U$ is unitary, it follows  that there is a basis for ${\mathcal C}$ in which
$U$ is diagonal. Let ${\mathcal B} = \{|s\rangle \}$ denote this basis. Let $\lambda_{s}$ denote the eigenvalue of $U$ corresponding to $|s\rangle$ so
that $U |s\rangle = \lambda_{s} |s\rangle.$ Let $\alpha_{s,s'}$ be the matrix element for $\partial$ so that
$$\partial |s\rangle = \sum_{s'}\alpha_{s,s'} |s' \rangle$$ where $s'$ runs over a set of basis elements so that $i(s') = i(s) + 1.$
\bigbreak

\noindent {\bf Lemma.} With the above conventions, we have that for $|s' \rangle$ a basis element such that $\alpha_{s,s'} \ne 0$ then
$\lambda_{s'} = - \lambda_{s}.$
\smallbreak

\noindent {\bf Proof.} Note that $$U \partial |s \rangle = U(\sum_{s'} \alpha_{s,s'} |s' \rangle) = \sum_{s'} \alpha_{s,s'} \lambda_{s'}|s' \rangle$$
while $$\partial U |s \rangle = \partial \lambda_{s} |s \rangle = \sum_{s'} \alpha_{s,s'} \lambda_{s}|s' \rangle.$$
Since $U \partial + \partial U = 0,$ the conclusion of the Lemma follows from the independence of the elements in the basis for the Hilbert space. //
\bigbreak

\noindent In  this way we see that eigenvalues will propagate forward from ${\mathcal C}^{0}$ with alternating signs according to the appearance
of successive basis elements in the boundary formulas for the chain complex. Various states of affairs are possible in general, with new eigenvaluues
starting at some ${\mathcal C}^{k}$ for $k > 0.$ The simplest state of affairs would be if all the possible eigenvalues (up to multiplication by $-1$) for $U$
occurred in ${\mathcal C}^{0}$ so that  $${\mathcal C}^{0} =\oplus_{\lambda}{\mathcal C}^{0}_{\lambda}$$ where $\lambda$ runs over all the distinct eigenvalues of $U$
restricted to ${\mathcal C}^{0},$
and ${\mathcal C}^{0}_{\lambda}$ is spanned by all $|s \rangle$ in ${\mathcal C}^{0}$ with $U |s \rangle = \lambda |s \rangle.$ Let us take the further
assumption that for each $\lambda$ as above, the subcomplexes
$${\mathcal C}^{\bullet}_{\lambda}: {\mathcal C}^{0}_{\lambda} \longrightarrow {\mathcal C}^{1}_{-\lambda} \longrightarrow {\mathcal C}^{2}_{+\lambda} \longrightarrow \cdots
{\mathcal C}^{n}_{(-1)^{n}\lambda}$$ have ${\mathcal C} = \oplus_{\lambda} {\mathcal C}^{\bullet}_{\lambda} $ as their direct sum. With this assumption about the chain
complex, define
$|\psi \rangle = \sum_{s} |s\rangle$ as before, with $| s \rangle$ running over the whole basis for ${\mathcal C}.$  Then it follows just as in the beginning of
this section that $$\langle \psi|U| \psi \rangle = \sum_{\lambda} \lambda \chi(H(C^{\bullet}_{\lambda})).$$ Here $\chi$ denotes the Euler characteristic
of the homology. The point is, that this formula for $\langle \psi|U| \psi \rangle$ takes exactly the form we had for the special case of Khovanov
homology (with eigenvalues $(-1)^{i}q^{j}$), but here the formula occurs just in terms of the eigenspace decomposition of the unitary transformation
$U$ in relation to the chain complex. Clearly there is more work to be done here and we will return to it in a subsequent paper.
\bigbreak

\noindent {\bf Remark on the density matrix.}
Given the state $| \psi \rangle,$ we can define the {\it density matrix} $$\rho = | \psi \rangle \langle \psi|.$$ With this definition it is immediate
that $$Tr(U \rho) = \langle \psi|U| \psi \rangle$$  where $Tr(M)$ denotes the trace of a matrix $M.$ Thus we can restate the form of our result about
Euler characteristics as $$Tr(U \rho) = \sum_{\lambda} \lambda \chi(H(C^{\bullet}_{\lambda})).$$ In searching for an interpretation of the Khovanov
complex in this quantum context it is useful to use this reformulation. For the bracket we have
$$\langle K \rangle = <\psi|U|\psi> = Tr(U \rho).$$
\bigbreak

The Hilbert space for expressing the bracket polynomial as a quantum statistical amplitude is quite naturally the chain complex for
Khovanov homology with complex coefficients, and the unitary transformation that is the structure of the bracket polynomial acts on the homology of this
chain complex. This means that the homology classes contain information preserved by the quantum process that underlies the bracket polynomial.
We would like to exploit this direct relationship between the quantum model and the Khovanov homology to obtain deeper information about the relationship of
topology and quantum information theory, and we would like to use this relationship to probe the properties of these topological invariants.
\bigbreak

\noindent {\bf Remark on calculation of homology}. In this section we have a Hilbert space
 $C(K)$ with a boundary operator $\partial:C(K) \longrightarrow C(K).$ We define (without specifying the 
 grading) the {\em homology} of $C(K)$ to be $$H(K) = Kernel(\partial)/Image(\partial).$$
 It is a natural to ask for a quantum algorithm to compute $H(K)$. We have not attempted to solve this question in the present paper. A good solution to this question would yield many results, and in our case it would give a quantum algorithm for computing Khovanov homology.
\bigbreak

\noindent {\bf Remark on quantum field theory.} In \cite{Witten,WittenB} Witten gives a quantum field theoretic interpretation of the Jones polynomial and, more recently, a quantum field theoretic interpretation of Khovanov homology. It is an open problem to compare the quantum context that this paper describes for Khovanov homology with the context constructed in Witten's work.
\bigbreak


\begin{thebibliography}{99}

\bibitem{Ah1}
D. Aharonov, V. Jones, Z. Landau,
A polynomial quantum algorithm for approximating the Jones polynomial,
quant-ph/0511096.

\bibitem{Ah2}
D. Aharonov, I. Arad,
The BQP-hardness of approximating the Jones polynomial,
quant-ph/0605181 

\bibitem{Baxter}  R.J. Baxter.  Exactly Solved Models in Statistical Mechanics.  Acad. Press (1982).

\bibitem{BK}
R. V. Buniy and T. W. Kephart, Glueballs and the universal energy spectrum of tight knots and links. Int.J.Mod.Phys. A20 (2005) 1252-1259. physics.hep-ph/0408027.

\bibitem{JO} 
V.F.R. Jones, A polynomial invariant for links via von Neumann algebras,
Bull. Amer. Math. Soc. {\bf 129} (1985), 103--112.

\bibitem{JO1}
 V.F.R.Jones.  Hecke algebra representations of braid groups and link polynomials.  Ann. of Math.  126 (1987), pp. 335-338.
 
\bibitem{JO2}
V.F.R.Jones.  On knot invariants related to some statistical mechanics models.  Pacific J. Math., vol. 137, no. 2 (1989), pp. 311-334.

\bibitem{KaB}  
L.H. Kauffman, State models and the Jones polynomial, Topology {\bf 26} (1987),
395--407.

 \bibitem{KA89}  
L.H. Kauffman,  Statistical mechanics and the Jones polynomial,  AMS
Contemp. Math. Series  {\bf 78} (1989), 263--297.

\bibitem{KL}
L.H. Kauffman and S. Lins, {\em Temperley-Lieb Recoupling Theory and Invariants of Three-Manifolds},
Princeton University Press, Annals Studies {\bf 114} (1994). 

\bibitem {KP}
L.H. Kauffman, {\em Knots and Physics}, World Scientific Publishers (1991), 
Second Edition (1993), Third Edition (2002).

\bibitem{QCJP}
L.H. Kauffman, Quantum computing and the
Jones polynomial, math.QA/0105255, in {\em Quantum Computation and Information}, S. Lomonaco, 
Jr. (ed.), AMS CONM/305, 2002, pp.~101--137.

\bibitem{KhoJones}
L. H. Kauffman. A quantum model for the Jones polynomial, Khovanov homology and generalized 
simplicial homology. in ``Cross disciplinary advances in quantum computing'', Contemp. Math.
Vol. 536, ed. by Mahdavi, Koslover and Brown. Amer. Math. Soc. 2011.

\bibitem{KauffTopQ}
L. H. Kaufman and S. J.  Lomonaco. Quantum Topology and quantum computing,
 in ``Mathematics of Quantum Computation and Quantum Technology" 
 edited by Goong Chen, L. H. Kauffman and S. J. Lomonaco, 
 Chapman and Hall/CRC Applied Mathematics and Nonlinear Science Series (2008), pp. 409-504.
 
\bibitem{KauffQKnots}
L.  H. Kauffman and S.  J. Lomonaco Jr., Quantum knots, in {\it 
Quantum Information and Computation II -- Proceedings of Spie, 12 -14 April 2004} (2004), 
ed. by Donkor Pirich and Brandt, Intl. Soc. Opt. Eng, pp. 268-284.

\bibitem{KauffQknots1} 
L. H.Kauffman and S. J. Lomonaco, Quantizing Knots, Groups and Graphs, 
to appear in the SPIE Proceedings, April 28-29, (2011).

\bibitem{LomQKnots}
 S. J. Lomonaco and L. H. Kauffman, Quantum Knots and Mosaics, 
Journal of Quantum Information Processing, Vol. 7, Nos. 2-3, (2008), pp. 85 - 115. 
Republished inAMS PSAPM/68, (2010),  177-208.
http://arxiv.org/abs/0805.0339

\bibitem{LomQknots1}
 S.  J. Lomonaco, L. H. Kauffman. Quantum Knots and Lattices, or a Blueprint for Quantum Systems that Do Rope Tricks. in ``Quantum Information Science and Its Contributions to Mathematics", Proc. Symposia in Appl. Math. Vol. 68,  ed. by S. J. Lomonaco, Amer. Math. Soc. (2010).arXiv:0910.5891.


\bibitem{LomQknots2} 
S. J. Lomonaco , and L. H.Kauffman, Quantizing Braids and Other Mathematical
Structures: The General Quantization procedure, 
to appear in the SPIE Proceedings, April 28-29, (2011).

\bibitem{Spin}
L. H. Kauffman and S. J. Lomonaco Jr.  $q$-deformed spin networks, knot polynomials and anyonic
topological quantum computation. J. Knot Theory Ramifications 16 (2007), no. 3, 267--332. 

\bibitem{Fibonacci}
L. H. Kauffman and S. J. Lomonaco Jr., The Fibonacci Model and the Temperley-Lieb Algebra.
{\it International J. Modern Phys. B}, Vol. 22, No. 29 (2008), 5065-5080.

\bibitem{Three}
L. H. Kauffman and S. J. Lomonaco, Jr., A 3-Stranded Quantum Algorithm for the Jones Polynomial,
Proc. SPIE, vol. 6573, (2007), 65730T-1-65730T-13. http://arxiv.org/abs/0706.0020. 

\bibitem{QM}
S. J. Lomonaco Jr. and L. H. Kauffman, A Quantum Manual for Computing the Jones Polynomial,  Proc. SPIE on Quantum Information and Computation VI, Vol. 6976, (2008), 
pp. 69760K-1 to 69760K-4.

\bibitem{Group1}
R. Marx, A. Fahmy, L. H. Kauffman, S. J. Lomonaco Jr., A Sp\"{o}rl, N. Pomplun, 
T. SchulteHerbr\"{u}ggen,
J. M. Myers, and S. J. Glaser, Nuclear-magnetic-resonance quantum calculations of the Jones polynomial,  {\it Physical Review A}, (2010).

\bibitem{NSSF}
C. Nayak, E. H. Simon, A. Stern, M. Freedman, S.  Das Sarma, Non-abelian anyons and topological quantum computation. Rev. Modern Phys. 80 (2008), no. 3, 1083--1159.

\bibitem{Rasetti}
M. Rasetti and T. Regge, Vortices in He II, current algebras and quantum knots. Physica {\bf 80A}
(1975) 217-233. North Holland Pub. Co.

\bibitem{Shor}
E. Farhi, D. Gosset, A. Hassidim, A. Lutomirski, P. Shor,
Quantum money from knots. arXiv:1004.5127 [quant-ph].

\bibitem {Viro} O. Viro (2004), Khovanov homology, its definitions and ramifications,
{\it Fund. Math.}, {\bf 184} (2004), pp. 317-342.

\bibitem{Witten} E. Witten. Quantum Field Theory and the Jones Polynomial. Comm. in Math. Phys.
Vol. 121 (1989), 351-399.

\bibitem{WittenB} E. Witten. Fivebranes and knots. arXiv:1101.3216 [math.GT].

\end{thebibliography}
\end{document}